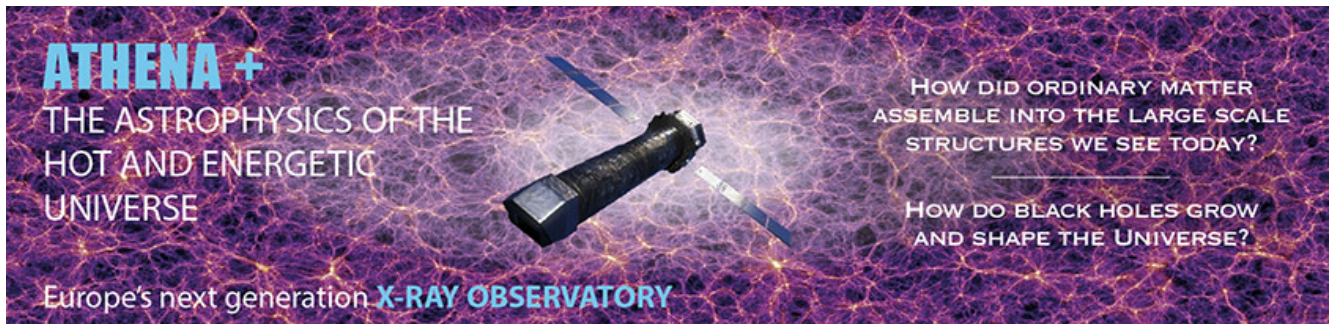

# The Hot and Energetic Universe

An *Athena+* supporting paper

# The Astrophysics of galaxy groups and clusters with *Athena+*

Authors and contributors

**S. Ettori, G. W. Pratt,** J. de Plaa, D. Eckert, J. Nevalainen, E.S. Battistelli, S. Borgani, J.H. Croston, A. Finoguenov, J. Kaastra, M. Gaspari, F. Gastaldello, M. Gitti, S. Molendi, E. Pointecouteau, T.J. Ponman, T.H. Reiprich, M. Roncarelli, M. Rossetti, J.S. Sanders, M. Sun, G. Trinchieri, F. Vazza, M. Arnaud, H. Böringher, F. Brighenti, H. Dahle, S. De Grandi, J.J. Mohr, A. Moretti, S. Schindler

# 1. EXECUTIVE SUMMARY

As the nodes of the cosmic web, clusters of galaxies trace the large-scale distribution of matter in the Universe. They are thus privileged sites in which to investigate the complex physics of structure formation. Over 80% of their total mass, which can reach more than $10^{14}$ $M_\odot$, is contained in the form of dark matter. The remaining mass is composed of baryons, most of which (about 85%) exists in the form of a diffuse, hot, metal-enriched plasma that radiates primarily in the X-ray band – the intracluster medium (ICM). The radiation from this gas and from the member galaxies allows us to observe and study the interplay between the dark matter and the hot and cold components of the baryonic mass budget. However, the complete story of how these structures grow, and how they dissipate the gravitational and non-thermal components of their energy budget over cosmic time, is still beyond our grasp. X-ray observations of the evolving cluster population provide a unique opportunity to address such fundamental open questions as:

- ■ • How do hot diffuse baryons accrete and dynamically evolve in dark matter potentials?
- ■ • How and when was the energy that we observe in the ICM generated and distributed?
- ■ • Where and when are heavy elements produced and how are they circulated?

**An X-ray observatory with large collecting area and an unprecedented combination of high spectral and angular resolution, such as *Athena+*, offers the *only* way to make major advances in answering these questions.** Such an observatory will permit a definitive understanding of the formation and evolution of large-scale cosmic structure through the study of the cluster population.

In this Supporting Paper we discuss the impact of *Athena+* on the study of galaxy group and cluster astrophysics. We focus on observations of nearby (z < 0.5) systems, where *Athena+* will revolutionize our understanding of the basic process of energy transfer into the ICM, transform our view of the outer regions of galaxy clusters where material continues to accrete, and allow us to track the generation and diffusion of metals in the intracluster gas.

# 2. THE THERMODYNAMIC PROPERTIES OF LARGE SCALE STRUCTURE

Cosmic structure formation and evolution are central issues in cosmology. A robust cosmological framework, showing that 95% of the total mass-energy of the Universe is contained in cold dark matter and dark energy, has recently been established[1]. The remaining 5% of baryonic, matter collects in the deep dark matter potential wells and forms galaxies and clusters of galaxies. However, we do not understand this process well enough to be able to link our observations to the underlying theoretical models, which hampers our ability to make precise predictions of what we see from first principles.

As the nodes of the web of cosmic structure, galaxy clusters are unparalleled sites at which to investigate the complex physics of structure formation. Over 80% of their total mass, which can range from a few times $10^{13}$ $M_\odot$ up to more than $10^{15}$ $M_\odot$, is contained in the form of dark matter that can at present only be indirectly detected. The remaining mass is composed of baryons, of which over 85% are contained in the Intra-cluster Medium, a diffuse, hot (T > $10^7$ K) plasma that radiates primarily in the X-ray band. X-ray observations are thus absolutely critical for the study of this majority baryonic component.

In the present structure formation scenario, clusters result from the gradual, hierarchical merging of smaller structures of dark and luminous matter accreted along filaments. A large amount of baryonic mass is expected to be found along the filamentary structure of the cosmic web, at the interface of the Warm-Hot Intergalactic Medium (WHIM; see Kaastra, Finoguenov et al. 2013, *Athena+* supporting paper) with the external regions of groups and clusters of galaxies. These cluster and group outskirts, lying across the virial radius and occupying about 85% of the cluster volume, are also expected to display strong energetic activity as material is accreted into the dark matter potential. They host the most intense events of the cosmic growth process associated with accretion shocks and the transfer of energy into the ICM through merging events. These in turn trigger large-scale motions and perturb various ICM equilibrium states such as hydrostatic equilibrium, thermal equilibrium and equipartition, and ionization equilibrium of the gas (Reiprich et al. 2013).

---

[1] See for example the recent study of the cosmic microwave background anisotropies with the ESA mission *Planck.*

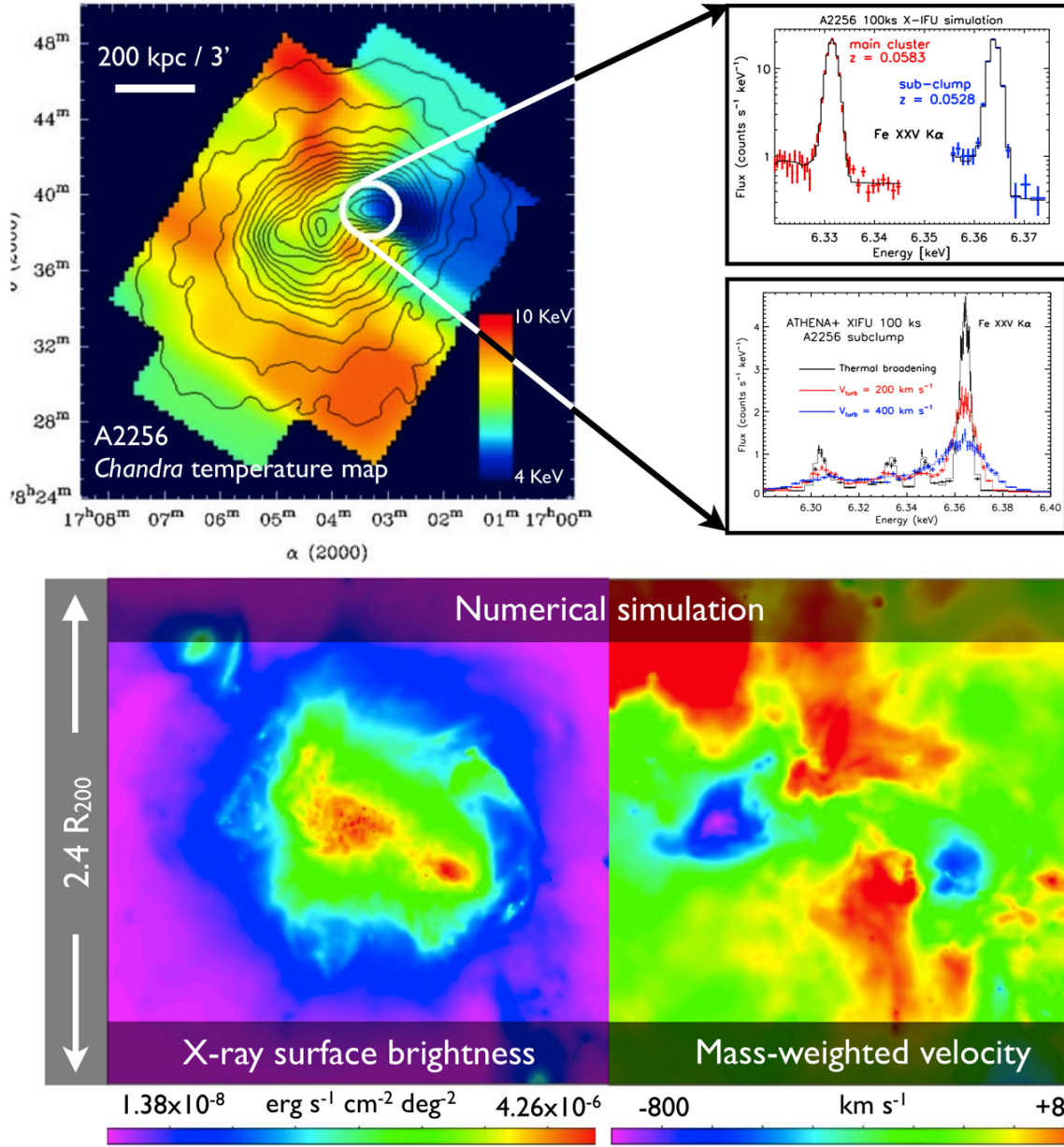

Figure 1: Upper panels: Bulk motion (the best-fit model -blue- of the emission line associated to the sub-clump has been scaled up for sake of clarity) and turbulent broadening of the Fe XXV Kα line over a circular region of 1.5 arcmin in radius centred on the sub-clump accreting onto the main body of A2256 ($z$=0.053), from a 100 k-sec simulation with Athena+ X-IFU. Simulated spectra convolved with turbulent velocities of 200 (red) and 400 (blue) km s$^{-1}$ are shown. For an input velocity of 0, 200, 400 km s$^{-1}$, the 1σ statistical uncertainty is +20, ±5, ±10 km s$^{-1}$, respectively (see also Nevalainen 2013). Residual bulk motions and gas turbulence add to the thermal pressure of the gas. Precise measure measurement of these quantities would cement our understanding of the violent process of structure formation. Lower panels: The high statistical precision of the redshift measurements of the Fe XXV K alpha lines with X-IFU would enable mapping of the radial components of the bulk motions due to recent mergers in nearby ($z$~0.1) clusters of galaxies for the first time, thus probing the process of structure formation in the act. At a spatial resolution of 0.1 R500, velocities of 10 km s$^{-1}$ can be detected at the cluster centre, and 50 km s$^{-1}$ at 0.25 R500 (3 σ c.l.). Single velocity measurements of ~1000 km s$^{-1}$ (3 σ c.l.) within the full X-IFU FOV at a distance of 0.5 R500 from the cluster centre can be achieved in a 50 ks off-axis pointing. X-IFU would enable the mapping of the turbulent velocities in nearby ($z$~0.1) clusters of galaxies. Velocities of 50 and 100 km s$^{-1}$ can be resolved at the cluster centre and at 0.25 R500 with a spatial resolution of 0.1 R500 (3 σ c.l.). At higher redshifts, using the emission in the full X-IFU FOV (i.e. assuming that all the ICM within the FOV has the same velocity) both turbulent and bulk motion velocities of 100 km s$^{-1}$ can be detected to $z$=0.7, while at $z$=1.2 a velocity level of 200 km s$^{-1}$ is detectable (3 σ c.l.). These estimates only include uncertainties on the statistical precision of the velocity shifts and line broadening measurements.

Galaxy clusters form from the most active portions of the high-redshift universe. Proto-cluster regions, which are currently mapped as concentrations of strongly star-forming galaxies at $z \sim$ 2-4 (e.g., Steidel et al. 2011), are characterised by intense star formation and associated galactic nuclear activity and metal enrichment of the high-redshift inter-galactic medium (IGM). Such intense activity in turn leaves its fingerprints in the thermodynamic and chemical properties of the low-redshift ICM. By studying the radiation from the hot gas and the galaxies in galaxy clusters, we can observe the interplay between the hot and cold components of the baryonic mass budget and the dark matter. X-ray observations with high spectral and angular resolution offer the *critical* way to make major advances in the understanding of the formation and evolution of large-scale cosmic structures through the study of the cluster population.

## 3. HOW DO HOT BARYONS DYNAMICALLY EVOLVE IN DARK MATTER POTENTIALS?

A complete story of structure formation from first principles is still beyond our grasp, due principally to our incomplete knowledge of cluster physics. For example, the gravitational energy released during cluster formation is expected to be dissipated in the intra-cluster gas, generating turbulence and producing kiloparsec-scale bulk motions, and yet this process has never been conclusively directly observed. Moreover, the contribution of the non-thermal component of the cluster energy budget over time, and its influence on the formation and intrinsic properties of galaxy groups and clusters, is basically unknown.

Current X-ray observations of the bright central regions indicate that many present epoch clusters are indeed not relaxed systems, but are scarred by shock fronts and contact discontinuities (Markevitch & Vikhlinin 2007), and that the fraction of unrelaxed clusters increase with redshift, as expected in a hierarchical scenario of structure formation. Although the gas evolves in concert with the dark matter potential, this gravitational assembly process is complex, as illustrated by the temporary separations of dark and X-ray luminous matter in massive merging clusters such as the "Bullet Cluster" (Clowe et al. 2006). At present, only upper limits on bulk motion velocities of the coolest, brightest, central gas are accessible (e.g., measurements with the Reflection Grating Spectrometer on board XMM-Newton indicate $v_{bulk} \leq$ 300-500 km s$^{-1}$, Sanders & Fabian 2013). A significant improvement in spectral resolution will be reached with upcoming *Astro-H* observations, but these will again be limited to the brightest regions. Further progress can only be achieved if we can truly map velocities and turbulence in the intra-cluster gas. This requires good angular resolution in combination with an improvement of more than an order of magnitude in collecting area compared to current instrumentation.

The true details of the on-going cluster assembly process, mainly traced by X-ray emission from a multi-temperature gas, will be resolved with *Athena+*. Accurate measurements of the distribution of the gas emission measures and of the centroids and widths of the ICM emission lines will be available with the high spectral resolution of the X-IFU X-ray calorimeter, coupled to the large collecting area of the X-ray mirrors. *Athena+* will bring exactly the key new observational capability - large collecting area, combined with high angular and spectral resolution - that will enable gas flow velocities and gas turbulence to be locally measured in clusters of galaxies using X-ray line width and position measurements. The *Athena+* X-IFU will allow such measurements to be achieved with an accuracy of around 10 – 20 km s$^{-1}$ for turbulent velocities of 100-1000 km s$^{-1}$ (see Fig. 1). This is sufficient to measure the thermal broadening in typical clusters, and hence will reveal any significant non-thermal pressure associated with gas motions, as well as bulk velocities. *Athena+* will thus allow the level of turbulence and the dynamics of the gas to be mapped routinely in the central regions (R ≤ 0.25 R$_{500}$)[2] of nearby (z ≤ 0.1) clusters, for the first time. Sub-cluster velocities and directions of motions will be measured by combining velocity shifts measured from X-ray spectra (which give relative line-of-sight velocities) and total sub-cluster velocities deduced from temperature and density jumps across merger shocks or cold fronts (Markevitch & Vikhlinin 2007). These measurements, combined with high quality lensing observations (from HST and future instruments such as, e.g., LSST and Euclid), will explain how the hot gas reacts to the evolving dark matter potential, and give a crucial insight into the dynamics of clusters in three dimensions. They will also reveal the role of shocks and turbulence in re-accelerating particles and generating the radio-emitting relics and haloes that are observed in some clusters (e.g. with LOFAR).

---

[2] R$_{500}$ is the radius within which the mean mass density is 500 times the critical density of the Universe at the cluster redshift. It corresponds approximately to 10 arcminutes for a 5 keV cluster at $z$ = 0.1.

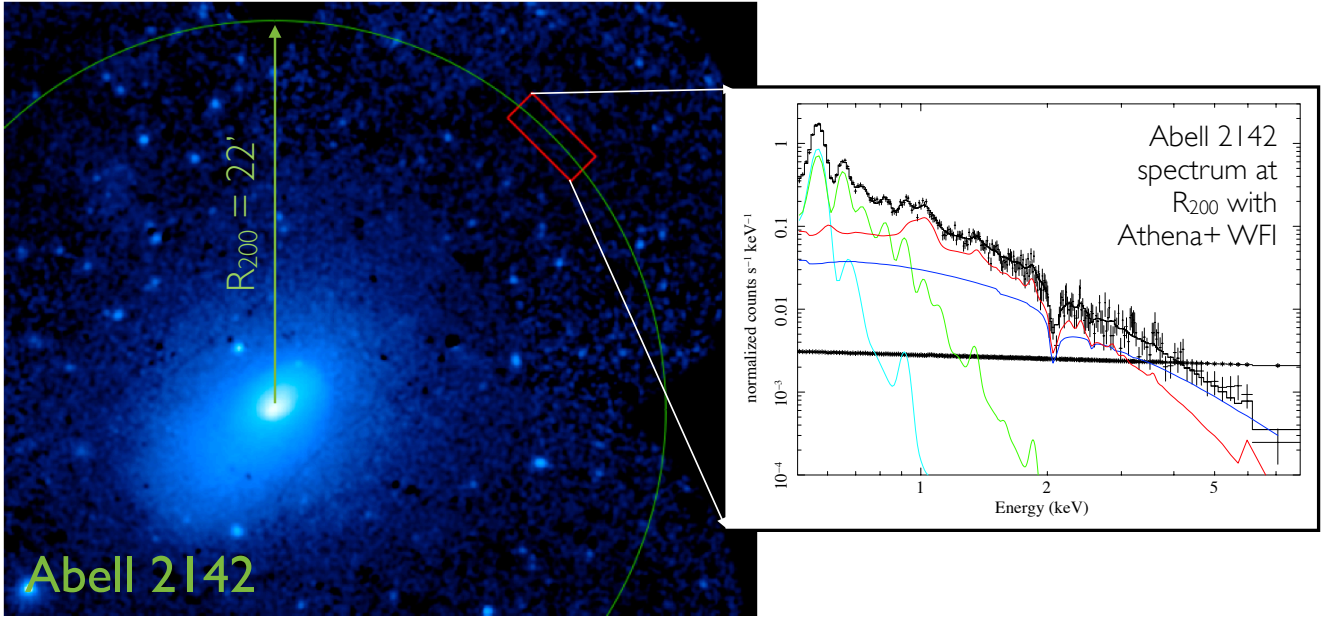

**Figure 2:** The combined X-IFU and WFI exposures will provide a temperature profile that will improve by an order of magnitude the present constraints on the thermodynamical properties of the ICM out to $R_{500}$ and beyond. This plot presents a XMM mosaic of A2142 at $z=0.09$ (*right*) and a 100 k-sec WFI exposure of a region of 10 arcmin$^2$ across the virial radius (*left*): the gas emissivity, temperature and metallicity are constrained with relative uncertainties of 2, 3 and 18 per cent, respectively, at 90% confidence (see also Appendix).

High-resolution spectroscopy from the *Athena+* X-IFU will resolve the emission complex around 6.7 keV arising from He-like Fe lines and their satellites, allowing us to use these lines as strong and independent temperature diagnostic of the ICM in the range $10^7$-$10^8$ K (Porquet et al. 2010). This measurement is particularly important to verify the standard assumption that the ICM is a tenuous gas in collisional ionisation equilibrium. This may not in fact be the case in the cluster outskirts, due to the low density and short timescales. In particular, the estimate of the intensity ratio between Fe XXV and Fe XXVI Kα line emissions trace the ionization equilibrium state that can be altered by ongoing mergers (e.g. Akahori & Yoshikawa 2010). In the shocked regions of the ICM, departures are also expected from the assumed Maxwellian electron distribution. In case of supra-thermal emission, this effect can be resolved via enhanced equivalent widths of the Fe XXV j-satellite (Kaastra et al. 2009).

**Through spatially resolved high-resolution spectroscopy, *Athena+* will measure intra-cluster gas motions and turbulence, showing how the baryonic gas evolves in the dark matter potential wells, and probe the true nature and physical state of the X-ray emitting plasma.**

## 4. HOW AND WHEN WAS THE ENERGY IN THE ICM GENERATED AND DISTRIBUTED?

Within the cluster atmosphere, it has been recognised that non-gravitational processes, particularly galaxy feedback from outflows created by supernovae (SN) and super-massive black holes, must play a fundamental role in the history of all massive galaxies as well as the evolution of groups and clusters as a whole. Feedback from AGN emission and stellar winds is likely (i) to produce the galaxy red sequence, (ii) to provide the extra energy required to keep large quantities of gas in cluster cores from cooling all the way down to molecular clouds, (iii) to account for the energy (i.e. entropy) excess observed in the central regions of the group and cluster ICM (e.g. Sun et al. 2009; Pratt et al. 2010). Understanding the process of energy injection, and whether the energy was introduced early in the formation of the first haloes (with further consequences for galaxy formation history), or gradually over time by AGN feedback, SN driven galactic winds, or an as-yet unknown physical process, is crucial to our understanding of structure formation and evolution. The specific case of AGN feedback in local ($z < 0.5$) systems is discussed further in Croston, Sanders et a. (2013, *Athena+* supporting paper) The various feedback processes, coupled with cooling, affect the overall thermodynamic properties of the ICM differently. The amount of energy modification, its localisation within the cluster volume, and the time-scale over which this modification occurs, all depend on the interplay between feedback and cooling. These processes also influence the production and circulation of the metals.

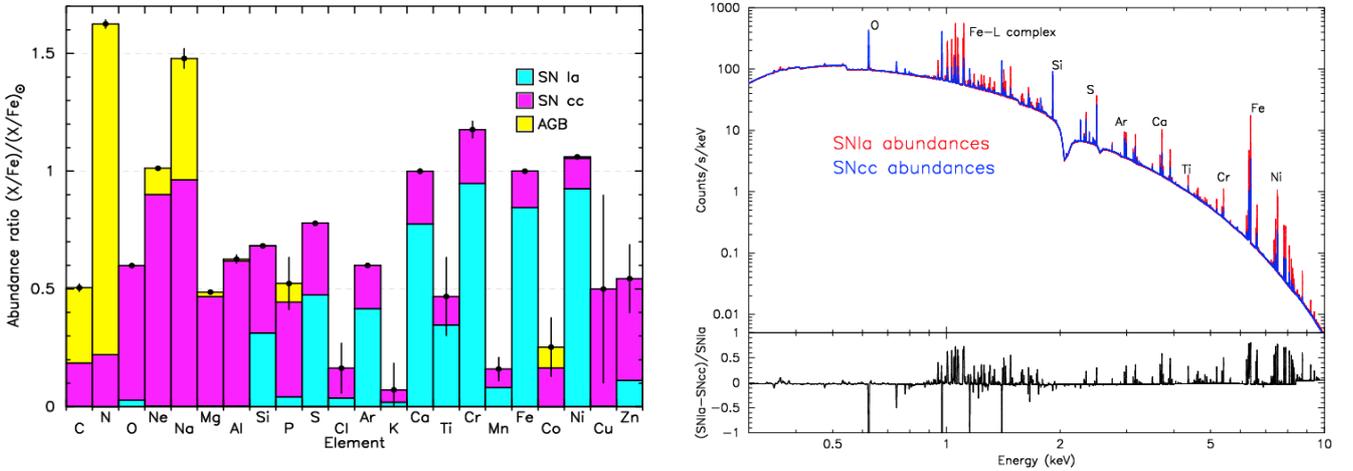

**Figure 3**: (*Left*) Expected abundance measurements with error bars for a typical cluster (AS1101) in a 100 ks X-IFU observation. The length of the colored bar shows the expected abundance ratios relative to solar. The colors indicate the fraction of the abundance produced by SNIa, SNcc and AGB stars for each individual element. Using the ensemble of cluster abundances measured with *Athena+*, constraints can be put on the SNIa explosion mechanism, the initial-mass function (IMF) of the stellar population and its initial metallicity. (*Right*) Two simulated X-IFU spectra of a typical 3 keV cluster at $z$=0.05. For the red spectrum, only type Ia products are shown with their expected abundance ratios. For the blue spectrum, only SNcc products are shown. It is clear from the residuals in the lower panel that SNIa and SNcc yields produce different line spectra. By fitting the true abundance measurements with a combination of SNIa and SNcc model yields, their relative contribution to the total yields can be estimated and constraints can be put on the IMF and the SNIa explosion mechanism.

The key to disentangling and understanding the respective role for each process lies in measurements of the gas entropy and metallicity (a direct probe of SN feedback). Entropy is generated by the shocks and gas compression during the hierarchical assembly process. It preserves a record of ICM cooling and heating because it always rises when energy is introduced and always falls when cooling carries energy away. The entropy generated by simple gravitational collapse is easy to predict (e.g., Voit et al 2005); deviations from this simple baseline model are due principally to the influence of non-gravitational processes linked to feedback and cooling. Current observations of local systems are limited to relatively bright objects and barely reach $R_{500}$ (e.g., Sun et al. 2009, Pratt et al. 2010). Other than in a few bright objects (e.g., Walker et al. 2012, Eckert et al. 2013) the outskirts are inaccessible. These regions are characterised by the ongoing accretion of material along large-scale filaments, and show the footprints of expanding shock waves triggered by internal merger activity. They thus provide crucial information on the most intense events of the cosmic growth process that shapes the entropy distribution. These processes also cause the distribution of the ICM in these regions to be clumpy and asymmetric, limiting the present observational constraints in reconstructing its physical properties.

*Athena+* will routinely allow spatially-resolved measurements of the entropy distribution out to large radial distances (at and beyond $R_{500}$), over all mass scales. This will help to localise the non-gravitational energy input and probe its effect over the entire halo volume from the centre to the outskirts (see Fig. 2 and A1), resolving where and how the accretion is taking place. Since non-gravitational effects are proportionally more effective in shallower potential wells, they are most noticeable in groups and poor clusters. In the present bottom-up hierarchical scenario these objects represent the building blocks of massive clusters and are the places where the majority of the galaxies (i.e., "cold" baryons) are thought to exist. With *Athena+*, observations of the entropy in the outskirts of nearby group-scale objects will be possible for the very first time. Spectra obtained with X-IFU and WFI will provide gas density and temperature profiles, and thus entropy and mass profiles, out to beyond $R_{500}$ for nearby group-scale objects ($M_{500} < 10^{14}$ $M_\odot$, or kT ~ 2 keV). In combination with observations of distant objects, these measurements will allow the sources of non-gravitational feedback, their influence within the cluster volume, their dependence on mass, and the timescales on which they operate, to be pinned down. The case of entropy and abundance evolution is discussed in more detail in Pointecouteau, Reiprich et al. (2013, *Athena+* supporting paper).

**We emphasise that no approved, or thus-far planned, X-ray mission will have the required characteristics for this sort of study. Only *Athena+* will be able to collect such information in the outskirts of nearby clusters and group-scale systems with moderate exposures of a few tens of kiloseconds, permitting resolution of the accreting region both spatially and spectroscopically. By making high-quality spatially resolved density and temperature measurements of groups and clusters over a large mass and redshift range, *Athena+* will reveal the epoch when their entropy excess was generated and its localisation within the halo volume.** The observations of the internal structure of the ICM described above will yield many new constraints on non-gravitational processes and their effect on the hot gas. However, an unambiguous picture can only be obtained in combination with the study of the integrated properties of the ICM such as luminosity, gas mass, temperature, etc. Non-gravitational effects change the expected normalisations, slopes, and dispersions of the relations between these integrated properties. However, present observational measurements are prone to strong selection biases, making it difficult to establish a robust observational baseline (e.g. Short et al. 2010, Reichert et al. 2011). In contrast, numerical simulations of cosmological structure formation, while at the point where modelling of all hydrodynamical and galaxy formation feedback processes is becoming feasible, are limited by our current understanding of the physics of the different feedback processes. Any advances in this field will be largely driven by observations, in combination with constant confrontation with cosmological numerical simulations.

Over the coming years, surveys such as eROSITA, LSST, Pan-STARRS, DES, and Euclid will deliver large, well-controlled samples of tens of thousands of clusters. High angular resolution *Athena+* observations of the ICM, the dominant baryonic mass component, in these systems, combined with radio observations (e.g. LOFAR, SKA), high angular resolution SZ millimetric data obtained from the new bolometric cameras optimised for large radio telescopes (e.g. GBT, IRAM, SRT, CCAT), observations of the cold baryons in galaxies (e.g. from JWST, ALMA, and E-ELT), and of the dark matter via gravitational lensing data (LSST, Pan-STARRS, Euclid) will provide, for the first time, the detail needed for a sufficiently critical comparison with theory. We expect that the major breakthrough of a detailed understanding of structure formation and evolution on cluster scales will come from simulation-assisted interpretation and modelling of these new generation observational data.

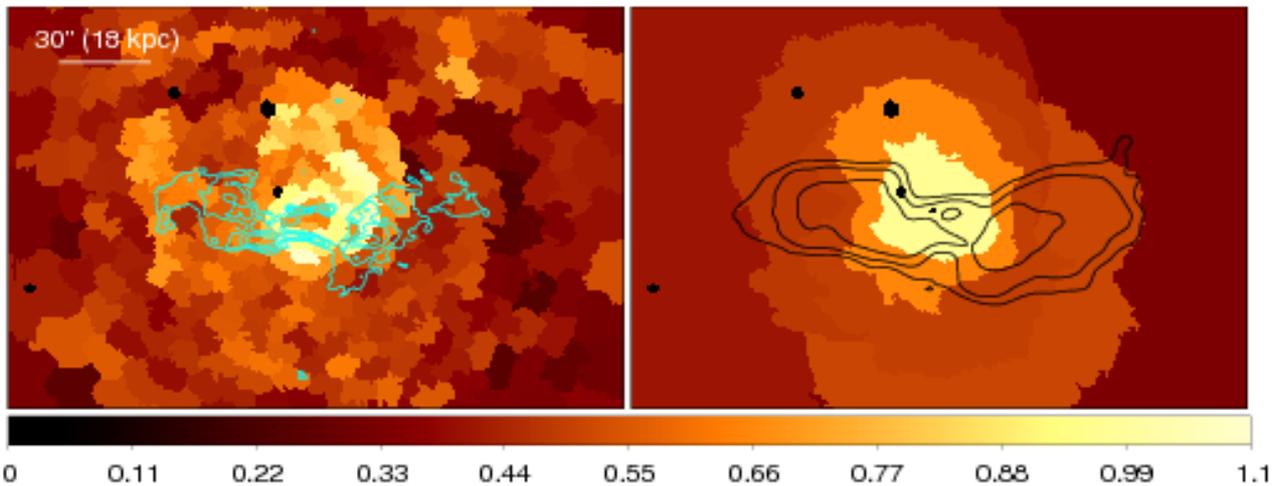

**Figure 4**: Metallicity maps recovered from the spectroscopic analysis of 20-ksec *Athena+* WFI (*left*) and *Chandra* (*right*), based on *Chandra* observations of Abell2199. This comparison of the effective spatial resolution achievable for metallicity studies demonstrates the *Athena+*'s ability to trace metal distributions on scales matched to jet and lobe structures in typical low-redshift clusters for the first time.

## 5. PRODUCTION AND CIRCULATION OF HEAVY ELEMENTS

Clusters are ideal laboratories to study the production and distribution of the metals (i.e. elements heavier than hydrogen, helium, lithium and beryllium). Their baryonic content has been continuously enriched with heavy elements since the first massive stars exploded as supernovae (SN). The abundance of metals in the ICM therefore corresponds to the time-integrated yield of the SN products released to the ICM from the member early-type galaxies (Arnaud et al. 1992). The cosmic history of Type Ia and core-collapse SN explosions is determined by the evolution and environmental dependence of the stellar initial mass function and star formation history in cluster galaxies. X-ray observations of the emission lines produced from the heavily ionized elements in the ICM are the only way to access information on their abundance and to probe their evolution to high redshift.

In local systems (z < 0.2), the abundance pattern of elements from O to the Fe group indicate that both type Ia and core-collapse SN contribute to the enrichment (de Plaa et al. 2007). **The large effective area and high spectral resolution of the *Athena+* X-IFU will allow the abundances of many common metals to be measured to exceptionally high precision.** For example, *Athena+* will help to constrain the main source of C and N, which can originate from a wide variety of sources including stellar mass loss from intermediate mass stars, and whose cosmic history is poorly known. Furthermore, measurements of trace elements such as Cr and Mn will be available for the first time (see Fig. 3 for the case of AS 1101). The abundance ratios of these elements provide additional constraints on the initial metallicity of the parent stellar populations. The Cr/Mn ratio is thought to depend on the initial metallicity of the SNIa progenitor system, while N and Na constrain the AGB star contribution. Abundance studies of a sample of 10-20 clusters with *Athena+* would be sufficient to give unprecedented insights into enrichment processes. Clusters with a high surface brightness distribution out to large radius (Perseus, M87, etc) would be privileged targets for this type of study.

A further fundamental question is how the metals produced in the galaxies are ejected and redistributed in the ICM. There are several mechanisms at work that can transport the metals from the Inter-Stellar Medium (ISM) of the galaxies, where these elements were produced, into the hot ICM. Feedback from AGN and superwinds from starbursts can drag metals from the galaxies into their environment. AGN jets appear to play an important role in uplifting enriched material to large radii (e.g. Simionescu et al. 2009), with heavy elements distributed anisotropically and aligned with the large-scale radio and cavity axes as is also suggested by simulations (e.g. Gaspari et al. 2011, Short et al. 2013). The radial ranges of the metal-enriched outflows are found to scale with jet power (e.g. Kirkpatrick et al. 2011). The heavy elements are transported beyond the extent of the inner cavities in all clusters, suggesting that this is a long-lasting effect sustained over multiple generations of outbursts. Detailed metal distribution mapping can also provide a powerful tracer of jet energy distribution, AGN-induced turbulence and gas motions (and the history of such motions; Rebusco et al 2005; see Fig.4), complementary to direct gas velocity measurements with X-IFU. Other processes such as ram-pressure stripping of in-falling galaxies, merger-induced sloshing, and galaxy-galaxy interactions can also release metals from the galaxies into the surrounding ICM. The radial distribution of metals in the ICM can give clues to the different processes at work. The relative abundances of different elements with radius provide also insight into the different enrichment mechanisms, their timescales, and how the gas is mixed by gas-dynamical processes. The current observational constraints are poor and controversial (Sato et al. 2007, Rasmussen & Ponman 2007, McCarthy et al. 2010). The unique capability for spatially resolved X-ray spectroscopy offered by *Athena+* X-IFU will capture these enrichment mechanisms in the act.

Some simulations show that the presence of AGN feedback makes radial metallicity profiles at low redshift shallower than those predicted from simulations with SN feedback alone (e.g. Fabjan et al. 2010). These shallower metallicity profiles are caused by the sudden expulsion of pre-enriched gas from groups and large galaxies at z ~ 2 - 3. However, to distinguish between models, abundance measurements are needed at radii R > $R_{500}$, which are inaccessible to current instruments that are limited to R < 0.5 $R_{500}$. The *Athena+* WFI will measure the most abundant elements out to unprecedented distances in the cluster atmosphere (i.e., up to and beyond $R_{500}$). This will show how elements are being distributed in the cluster volume, from the core to the outskirts of local systems. Studies of a few tens of systems with **Athena+ would allow us to establish what is happening to SNIa vs SNcc products as a function of position in a wide range of systems, which in turn would help to constrain feedback and mixing models. Together with measurements of more distant systems (see Pointecouteau, Reiprich et al. 2013, Athena+ supporting paper), this will place Athena+ in the unique position of providing a coherent picture of production and circulation of metals over a large portion of cosmic history.**

## 6. CONCLUDING REMARKS

An understanding of the physical processes involved in galaxy cluster assembly is an essential requirement for realistic models of cosmic structure formation. However, current X-ray data are unable to constrain the bulk motions of the ICM, cannot detect the hot gas much beyond $R_{500}$ (or only 10% of the cluster volume) even in bright local systems, and lack spatial resolution, throughput, and spectral resolution to map the distribution of the heavy elements within the cluster atmosphere beyond the bright central regions of the nearest objects. Overall, our vision of the thermodynamics and enrichment of the ICM should draw a consistent picture of the history of circulation of baryons through cosmic time. A major advance in X-ray sensitivity and spectral resolution, in combination with high spatial resolution, is needed to be able to make breakthroughs in these aspects of cluster physics. A large collecting area X-ray telescope such as *Athena+* will allow such progress to be made. It will also provide valuable insights into to the understanding of complementary datasets from ongoing and future radio observations (e.g. SKA), high angular resolution SZ millimetric data (e.g. GBT, IRAM, SRT, CCAT), observations of the cold baryons in galaxies (e.g. from JWST, ALMA, and E-ELT), and of the dark matter via lensing data (LSST, Pan-STARRS, Euclid).

## A. APPENDIX

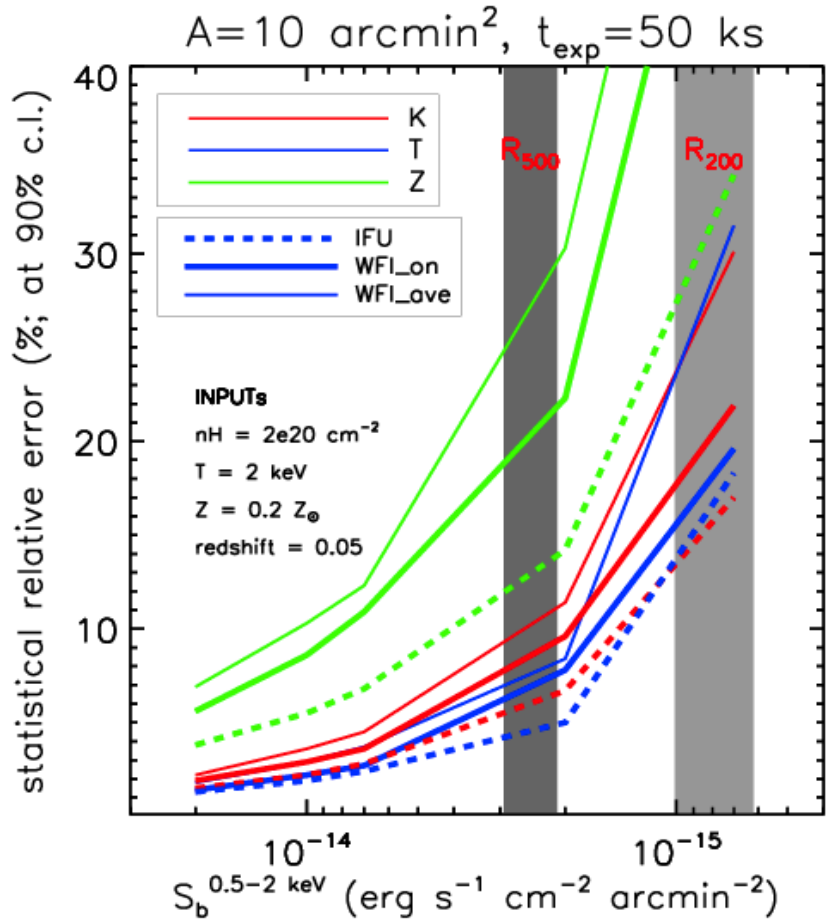

**Figure A1:** WFI & X-IFU constraints on the estimated properties (normalization K of a thermal spectrum modeled with apec, gas temperature T and metallicity Z) of the cluster outskirts as function of the X-ray surface brightness. Shaded regions show the expected location of $R_{500}$ and $R_{200}$. These spectral simulations include some randomisation of the known components of the X-ray total background (random fluctuations of 5% on two thermal components of the local background –modeled as in McCammon et al. 2002- and on cosmic background resolved by 80%; 1% random fluctuations on the galactic absorption $n_H$ and the particle background are considered).